\documentclass[%
 amsmath,amssymb,
 preprint,
 showkeys,
]{revtex4-1}

\usepackage{xcolor} \usepackage{ulem} \usepackage{changebar}

\usepackage{graphicx}
\usepackage{dcolumn}
\usepackage{bm}
\usepackage{tikz}
\usetikzlibrary{arrows}
\usepackage{ragged2e}
 \usepackage[colorlinks=true]{hyperref}
\hypersetup{urlcolor=blue, citecolor=blue, linkcolor=blue}

\begin{document}

\title{Chaotic bursting in semiconductor lasers}

\author{Stefan Ruschel}
\email{ruschel@math.tu-berlin.de}
\affiliation{%
Institute of Mathematics, Technical University of Berlin, Berlin, Germany
}%

\author{Serhiy Yanchuk}
\email{yanchuk@math.tu-berlin.de}
\affiliation{%
Institute of Mathematics, Technical University of Berlin, Berlin, Germany
}%


\begin{abstract}
We investigate the dynamic mechanisms for low frequency fluctuations
in semiconductor lasers subject to delayed optical feedback, using
the Lang-Kobayashi model. This system of delay differential equations
displays pronounced envelope dynamics, ranging from erratic, so called
low frequency fluctuations to regular pulse packages, if the time
scales of fast oscillations and envelope dynamics are well separated.
We investigate the parameter regions where low frequency fluctuations
occur and compute their Lyapunov spectrum. Using geometric singular
perturbation theory, we study this intermittent chaotic behavior and
characterize these solutions as bursting slow-fast oscillations. 
\end{abstract}


\maketitle


\section{Introduction}
In this paper, we theoretically investigate the phenomenon of low
frequency fluctuations (LFF) \cite{Morikawa1976,Risch1977} in a semiconductor
laser subject to delayed optical self-feedback. In the LFF regime,
the laser electric field amplitude is bound by a low-frequency envelope
that possesses amplitude dropouts at irregular times, i.e. the laser
flickers. If these LFFs are approximately periodic, they are also
referred to as regular pulse packages \cite{Heil2001,Davidchack2001}(RPP).
To model the dynamics of semiconductor lasers with feedback, Lang
and Kobayashi \cite{Lang1980}\,(LK) proposed the following system
\begin{eqnarray}
E^{\prime}(t) & = & \left(1+i\alpha\right)N(t)E(t)+\eta e^{i\phi}E(t-\tau),\label{eq:Edef}\\
N^{\prime}(t) & = & \varepsilon\left[J-N(t)-\left(1+2N(t)\right)|E(t)|^{2}\right].\label{eq:Ndef}
\end{eqnarray}
 Equations \eqref{eq:Edef}\textendash \eqref{eq:Ndef} model the
time-evolution of the laser in-cavity complex electric field $E(t)$
and excess carrier density $N(t)$. The parameter $\alpha$ is laser-specific,
the so called linewidth enhancement factor, $\eta>0$ and $\phi\in\mathbb{R}$
are the strength and phase shift of the feedback electric field. The
time-delay $\tau>0$ is the external cavity roundtrip time given in
units of the photon lifetime, i.e. $\tau=T/\tau_{p}$ where $T$ is
the roundtrip time and $\tau_{p}$ is the photon lifetime. The parameter
$\varepsilon>0$ is the ratio of the photon and carrier lifetimes
$\tau_{p}/\tau_{c}$, and it is usually a small parameter. $J$ is
the excess pump current, i.e. without self-feedback, the solitary
laser amplifies light when $J>0$.

Equations \eqref{eq:Edef}\textendash \eqref{eq:Ndef} are equivariant
with respect to optical phase-shifts $\left(E,N\right)\rightarrow\left(e^{i\psi}E,N\right),\,\psi\in\mathbb{R}$.
Therefore, they generically possess periodic solutions of the form
$E(t)=re^{i\omega t},$ $N(t)=n,$ such that $r>0$ and $\omega,n\in\mathbb{R}$
satisfy 
\begin{eqnarray}
r & = & \sqrt{\frac{J-n}{1+2n}},\\
i\omega & = & \left(1+i\alpha\right)n+\eta e^{i\left(\phi-\tau\omega\right)},\label{eq:Efasthopf}
\end{eqnarray}
referred to as External Cavity Modes (ECMs). 

ECMs are shown to play an important role in shaping the dynamics of
the LK system. In particular, a route to chaos has been shown via
a period doubling cascade of ECMs as $\eta$ is increased, where the
onset of chaos is related to the appearance of LFFs  \cite{Ye1993,Hohl1999}
and the chaotic attractor coexists with a stable ECM \cite{Mork1992,Levine1995},
the so-called maximum gain mode. It has been shown \cite{Lenstra1991,Sano1994,VanTartwijk1995,Fischer1996}
that LFFs can be considered as a chaotic itinerancy process on this
attractor: An LFF solution ''hops'' from one mode to another towards
the maximum gain mode, until it gets close to the stable manifold
of an ECM of saddle-type causing the LFF dropout event \cite{Lenstra1991,Sano1994,VanTartwijk1995,Fischer1996}.
This mechanism in the LK model agrees with the experiments \cite{Fischer1996,Torcini2006a}.
In view of the importance of the ECMs, their stability properties
have been studied in detail in Refs.~\onlinecite{Wolfrum2002,Rottschafer2007,Yanchuk2010}.
Further increasing the feedback strength however, leads to extensive
chaos, the so-called coherence collapse, without notable envelope
dynamics \cite{Lenstra1985,Heil1999}.

The structure of the paper is as follows. In Section~\ref{sec:props},
we determine parameter values in the $(\varepsilon,\tau)$-parameter plane,
where LFFs are observed numerically. Further, we study Lyapunov exponents \cite{Farmer1982}\,(LE)
of LFFs to quantify their chaotic behavior. In the particular case
of large delay, it is useful to consider asymptotic properties of
the Lyapunov spectrum leading to the distinction of weak versus strong
chaos \cite{Heiligenthal2011,Heiligenthal2013,DHuys2014,Jungling2015a,YanchukGiacomelli2017}.
For a superthreshold pump current, we show that the LFF corresponds
to a weakly chaotic orbit exhibiting short events of intermittent
strong chaos.

In the second part of the paper, Sec.~\ref{sub:Slowfast}, we provide
a complementary, geometric description of LFFs. For this, we discuss
stability properties of the off-state $E(t)=0$ and provide a simple
geometric viewpoint of the underlying dynamics for small $\varepsilon$.
In particular, we characterize LFFs as slow-fast oscillations, which
can be decomposed into pieces, each obtained from multiple time scale
analysis as $\varepsilon\to0$. We propose an averaged system of ODEs
describing the fast oscillations between the dropout events. In addition, we describe
the fast time-scale oscillations as interactions of eigen-modes of
the off-state. As this phenomenon bares some resemblance to a class
of slow-fast oscillations in neuronal systems modeled by ODEs \cite{Izhikevich2000},
we follow in notation and refer to them as bursting solutions.

\section{Parameter values where LFFs occur and properties of LFFs\label{sec:props}}

\subsection{Parameter region of LFFs\label{sec:detect}}

In this section, we numerically explore the parameter range where
LFFs can be observed. Since it is not feasible to consider all five
parameters in a numerical study, we fix $\alpha=5$, $\eta=0.1$,
and $\varphi=0.5$ (see Ref.~\onlinecite{Torcini2006a}) and vary
the parameters $\tau$, $\varepsilon$, and $J$. Note that the value
of the feedback phase $\varphi$ does not play an essential role for
the long-delay case, therefore, the observed results do not depend
on the chosen value of $\varphi$, at least for large feedback delay
$\tau$. 

\begin{figure}[t]
\noindent %
\begin{minipage}[t]{.5\linewidth}%
\includegraphics[width=1\linewidth]{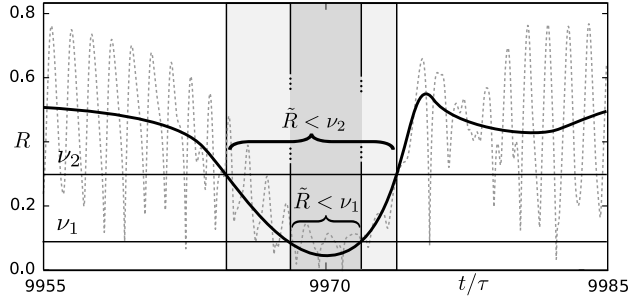}%
\end{minipage}\caption{Visual guidance for the computation of the measure $p$. The figure shows the amplitude $R(t)$ (gray, dotted) of a solution segment of Eqs.~\eqref{eq:Edef}--\eqref{eq:Ndef} for  $9955<t/\tau<9985$ and the parameters $\alpha=5$, $\eta=0.1$, $\phi=0.5$, $\varepsilon=0.003$, and $\tau=100$. In a dropout event, the time-$\tau$ averaged amplitude $\tilde{R}(t)$ (black, solid) falls below the threshold values $\nu_1=0.1$, $\nu_2=0.3$, respectively. The exact definition of $p$ (and $\tilde{R}$) is given in the text.}
\label{fig:p-def} 
\end{figure}

In order to quantify the dropout events, we introduce the probability
$p$ for the solution to exhibit an amplitude dropout using the following
empirical algorithm:\\
 1. For a given solution, we consider the running average $\tilde{R}\left(t\right)$
of the laser field amplitude $\tilde{R}(t)=\int_{t-\tau}^{t}|E(\theta)|d\theta$
over a window of length $\tau$.\\
 2. For a sufficiently long trajectory (here, we choose $10^{5}$
delay intervals, after some transient), we denote $\mathbb{P}(\{\tilde{R}\left(t\right)<\nu\})$
the probability of the laser amplitude to be below the threshold $\nu$.\\
 3. We compute the conditional probability 
\[
p_{\nu_{1},\nu_{2}}:=\mathbb{P}_{\left\{ \tilde{R}\left(t\right)<\nu_{2}\right\} }\left(\left\{ \tilde{R}\left(t\right)<\nu_{1}\right\} \right)=\frac{\mathbb{P}\left(\left\{ \tilde{R}\left(t\right)<\nu_{1}\right\} \right)}{\mathbb{P}\left(\left\{ \tilde{R}\left(t\right)<\nu_{2}\right\} \right)}
\]
for $\nu_{1}<\nu_{2}$, that is the fraction of time intervals, for
which the field amplitude satisfies $\tilde{R}\left(t\right)<\nu_{1}$
provided it is below $\nu_{2}$. Intuitively speaking, $\nu_{1}$ determines the maximum value of $\tilde{R}(t)$, which we tolerate within a dropout event and $\nu_{2}$ specifies the minimum value of $\tilde{R}(t)$, we require outside of a dropout event. The optimal choice of these threshold values
is empiric. By experimenting and testing the known LFF
regimes, we set the values to $\nu_{1}=0.1$ and $\nu_{2}=0.3$ (see Fig.~\ref{fig:p-def}). \\
4. Finally, we would like to discard the solutions that are oscillating
around the zero-amplitude threshold, since they are clearly not LFFs.
Therefore, we define the measure 
\[
p:=p_{\nu_{1},\nu_{2}}\frac{N_{\nu_{2}}}{N_{\nu_{1}}},
\]
where $N_{\nu}$ is the number of crossings of the corresponding threshold
value $\nu$. In the case when the solution possesses fast oscillations
around $\nu_{1}$ within one dropout event, the value of $N_{\nu_{1}}/N_{\nu_{2}}$
is large and, hence, the measure $p$ attains small values.

It is clear that such a defined quantity $p$ is largest if all dropouts
below the threshold $\nu_{2}$ attain almost only values smaller than
$\nu_{1}$. Hence, the higher values of $p$ are indicators of an
orbit to posses a dropout event, and, therefore, it is a candidate
for LFFs or RPPs. Here we should note that the values of $p$ close
to 1 can also indicate a high degree of regularity or simply convergence
to $E(t)=0$. For instance, for a rectangular piecewise constant (square
wave) function with two values $\tilde{R}_{1},\tilde{R}_{2}$, $\tilde{R}_{1}<\nu_{1}<\nu_{2}<\tilde{R}_{2}$,
the value of $p$ equals $1$. However, these parameter regimes have
been avoided carefully.

\begin{figure}[t]
\noindent %
\begin{minipage}[t]{.5\linewidth}%
\includegraphics[width=1\linewidth]{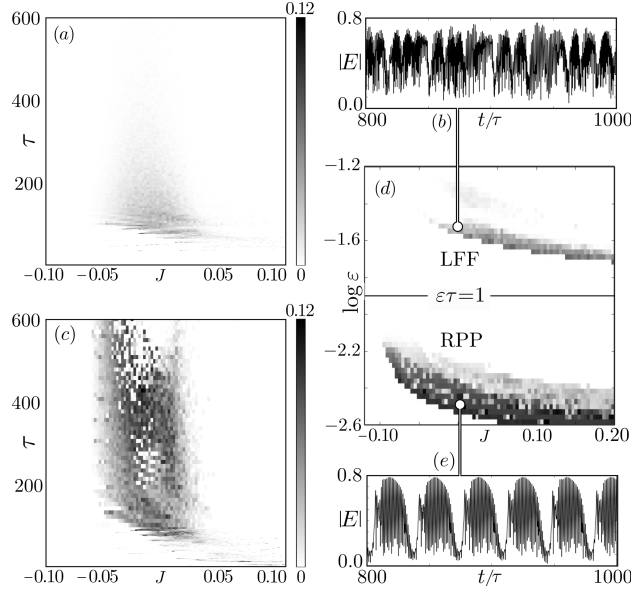}%
\end{minipage}\caption{Parameter ranges, where LFFs and RPPs are observed. Parameter $(J,\tau)$-diagram
in panel (a) shows the probability $p$ for the occurrence of LFFs
(see the description of $p$ in the text) for all other fixed parameters
$\varepsilon=0.03$, $\eta=0.1$, $\alpha=5$. Panel (c) shows the
same as (a) but with variable $\varepsilon=2.6/\tau$. One observes
that LFFs are more abundant in (c), where $\varepsilon$ is scaled
with $\tau$. Panel (d) shows the same LFF indicator in the parameter
plane $(J,\log\varepsilon)$ with fixed $\tau=85$. Two distinct parameter
regions are observed: one for LFFs and another for RPPs existing
for significantly smaller values of $\varepsilon$, such that $\varepsilon\tau<1$.
Panels (b) and (e) illustrate trajectories from the corresponding
parameter regions of (d). }
\label{fig:LFFdect} 
\end{figure}

Using the introduced quantity $p$, Fig.~\ref{fig:LFFdect} shows
the regions for the existence of LFFs with respect to parameters $J$,
$\varepsilon$, and $\tau$. In Fig.~\ref{fig:LFFdect}(a), we fix
parameters $\varepsilon=0.03,~\eta=0.1,~\alpha=5$, for which LFFs
have been reported experimentally in Ref.~\onlinecite{Torcini2006a}.
For this choice of parameters, LFFs do not appear for small values
of the time-delay, $\tau<40$. Moreover, LFFs cease to exist for
large values of $\tau$. In the $(J,\tau)$ parameter plane, $p$
attains its maximum for $J\approx0$ and $\tau\approx85$. The corresponding
solution for maximal $p$ is shown in Fig.~\ref{fig:LFFdect}(b)
and displays pronounced LFFs.

In order to investigate the interplay of $\varepsilon$ and $\tau$,
we fix the maximum-$p$ delay value $\tau=85$ and investigate the
influence of the parameter $\varepsilon\tau$ on the occurrence of
LFFs. Note that $\varepsilon\tau=T/\tau_{c}$ is the external cavity
roundtrip time given in units of the carrier lifetime. For the maximum-$p$
solution shown in Fig.~\ref{fig:LFFdect}(b), we have $\varepsilon\tau\approx2.6$.
We fix this relation when changing $\tau$ in Fig.~\ref{fig:LFFdect}(c).
More specifically, Fig.~\ref{fig:LFFdect}(c) shows the same as (a)
but with variable $\varepsilon=2.6/\tau$, as we vary the delay. One
observes clearly that LFFs are more abundant in Fig.~\ref{fig:LFFdect}(c).

Figure~\ref{fig:LFFdect}(d) shows the value of $p$ in the parameter
plane $(J,\log\varepsilon)$ with fixed $\tau=85$. We observe two
distinct parameter regions, where $p$ attains high values. In the
dark region located in the upper part of Fig.~\ref{fig:LFFdect}(d),
$\varepsilon\tau\in[1.7,2.7]$, we observe LFFs. The lower parameter
region in Fig.~\ref{fig:LFFdect}(d) corresponds to RPP solutions
(Fig.~\ref{fig:LFFdect}(e)), which have been reported in Refs.~\onlinecite{Heil2001,Davidchack2001}.
It is characterized by the values $\varepsilon\ll1$ and $\varepsilon\tau<1$.

Summarizing, Fig.~\ref{fig:LFFdect} shows parameter regions where
LFFs and RPPs are observed. These numerical results indicate the importance
of $\varepsilon\tau$ as a scaling parameter. Moreover, we observe
a clear distinction between LFFs and RPPs, namely, the LFFs are observed
for $\varepsilon\tau>1$ and RPPs for $\varepsilon\tau<1$ with a
clear gap between them, at least when the other parameters $\eta,\phi,\alpha$
are chosen as mentioned above.

\subsection{Lyapunov Spectrum\label{sec:LE}}

In this section, we provide a detailed analysis of the Lyapunov Spectrum
of LFFs. We briefly introduce some necessary concepts. Let $\left(E\left(t\right),N\left(t\right)\right)$
be a solution to Eqs.~\eqref{eq:Edef}\textendash \eqref{eq:Ndef}
and consider $\xi\left(t\right)$ the solution of the linearized equation
\[
\xi^{\prime}\left(t\right)=A(t)\xi\left(t\right)+B(t)\xi\left(t-\tau\right)
\]
 along the solution $\left(E\left(t\right),N\left(t\right)\right)$.
Here $f(\cdot)$ is the right-hand side of Eqs.~\eqref{eq:Edef}\textendash \eqref{eq:Ndef},
and $A(t):=Df\left(E\left(t\right),N\left(t\right),E(t-\tau)\right)$
and $B(t):=D_{\tau}f\left(E\left(t\right),N\left(t\right),E(t-\tau)\right)$
denote the Jacobians with respect to non-delayed and delayed arguments,
respectively. The Lyapunov Exponent \cite{Farmer1982}(LE) is defined~as 
\[
\lambda=\limsup_{t\to\infty}\frac{1}{t}\log\|\xi\left(t\right)\|.
\]
 It is convenient to consider a related concept, the so called instantaneous
or sub-LE, which is defined as 
\[
\sigma=\limsup_{t\to\infty}\frac{1}{t}\log\|\zeta\left(t\right)\|,
\]
where $\zeta\left(t\right)$ is the solution to the truncated linearized
equation $\zeta^{\prime}\left(t\right)=A(t)\zeta\left(t\right).$
The instantaneous LE can be used to distinguish the two distinct chaotic
regimes of weak and strong chaos \cite{Heiligenthal2011,Heiligenthal2013,Jungling2015a,YanchukGiacomelli2017}.
In particular, we call strong chaos, when the instantaneous LE is
positive, while weak chaos is characterized by negative instantaneous
LE and positive LE.

\begin{figure}
\noindent %
\begin{minipage}[t]{.5\linewidth}%
\includegraphics[width=1\linewidth]{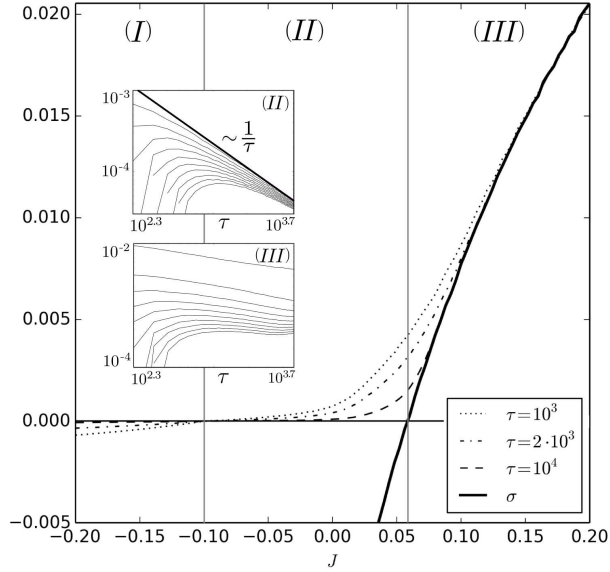}%
\end{minipage}\caption{Scaling behavior of Lyapunov Exponents (LE) for fixed values $\varepsilon=0.03,$
$\eta=0.1,$ and $\alpha=5$. The figure displays the largest LE $\lambda_{\max}$
for different values of the pump current $J\in[-0.2.0.2]$ and three
fixed values of the time-delay $\tau\in\left\{ 10^{3},2\cdot10^{3},10^{4}\right\} $.
The respective largest instantaneous LEs $\sigma$ coincide for all
considered values of $\tau$. We obtain 3 distinct regimes. (I): convergence
to equilibrium, (II): weak chaos ($\lambda=\mathcal{O}\left(1/\tau\right)$
as $\tau\to\infty$), (III): strong chaos ($\lambda=\mathcal{O}\left(1\right)$
as $\tau\to\infty$). To visualize the scaling behavior of $\lambda_{\max}$,
the 10 largest LE for $J=-0.05$ (II) and $J=0.07$ (III) are provided
in the insets in log-log plot.}
\label{fig:LE} 
\end{figure}

We have computed the largest LE and the instantaneous LE of solutions
to Eqs. \eqref{eq:Edef}\textendash \eqref{eq:Ndef} for varying values
of the pump current $J$ and different values of delay $\tau$ in
Fig.~\ref{fig:LE}. From left to right, we observe a convergence
to the off-state ($J<-\eta=-0.1$), weak chaos ($-0.1<J<0.06$), and
strong chaos ($J>0.06$). In the regime of weak chaos we observe $\lambda\sim1/\tau$.
Indeed, we have computed the ten largest LEs for fixed $J=-0.05$
for varying delay time $\tau$ in Fig.~\ref{fig:LE}(inset II) to
show this scaling property. Contrarily, in the case of strong chaos,
$\lambda_{\max}\to\sigma$ as $\tau\to\infty$, where $\lambda_{\max}$
is the largest LE. Fig.~\ref{fig:LE} shows that for $J=0.07$ the
largest LE remains at constant positive values as the delay increases. 

When comparing the domain of existence of LFFs in Fig.~\ref{fig:LFFdect}(a)
with Fig.~\ref{fig:LE} (both figures computed for $\varepsilon=0.03$),
we see that LFFs occur in the regime of weak chaos, at least those
shown in Fig.~\ref{fig:LFFdect}(a). However, this observation should
be considered with caution, since LFFs cease to exist for fixed $\varepsilon$ and very large $\tau$, as follows from Sec.~\ref{sec:detect}. Hence, one cannot
identify an appropriate LFF-trajectory asymptotically for $\tau\to\infty$
(and all other parameters fixed), as one would need for weak chaos.
Nevertheless, an approximate description can be done still for the
reported values of $\tau$ that are of order $10^{3}$. 

\begin{figure}
\includegraphics[width=.5\linewidth]{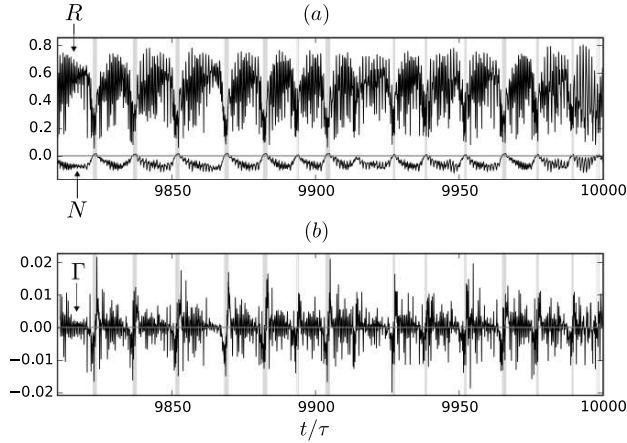}\caption{Panel (a): Time series of amplitude $R(t)=|E(t)|$ and carrier density
$N(t)$ of a solution to LK for fixed parameters $\alpha=5,\eta=0.1,\phi=0.5,\varepsilon=0.025,$
and $J=0.02.$ The intervals, where $N(t)\geq0$, are highlighted
by vertical gray lines with the width corresponding to the duration
of the event. These intervals clearly indicate the LFF dropout event.
Panel (b): Exponential rate of change of the amplitude $\Gamma(t)=\log(R(t)/R(t-\tau))/\tau$
from one delay interval to the next. LFFs dropout events are characterized
by larger values of $\Gamma(t)$, as the solution changes fast between
delay intervals. \label{fig:LFF-example}}
\end{figure}

A better understanding of the finite-time LEs can be achieved by considering
$E\left(t\right)$ as a time series given by the solution to the nonautonomous
system \eqref{eq:Edef}. Then, the instantaneous LE of it, can be
obtained from the truncated system 
\begin{equation}
E'(t)=(1+i\alpha)N(t)E(t),\label{eq:LET}
\end{equation}
where $N(t)$ is a solution of the full LK system. Equation \eqref{eq:LET}
is linear in $E$, hence, its instantaneous LE $\sigma_{E}$ is given by 
\begin{equation}
\sigma_{E}=\limsup_{t\to\infty}\frac{1}{t}\int_{0}^{t}N\left(s\right)\text{d}s.\label{eq:ILEalternat}
\end{equation}
In particular, this simple computation
reveals that the onset of strong chaos is due to positive values of
$N\left(t\right)$. In particular, if $N\left(t\right)\geq0$ for
some time-interval, it gives rise to a positive instantaneous finite
time LE, that is the instantaneous local rate of the growth at time $t$: $\sigma_{E,loc}(t)=\frac{1}{l}\int_{t-l}^{t}N\left(s\right)\text{d}s$
for certain $t$ and $l$. In the case $J>0$ and considering an LFF
trajectory, the positive values of $N(t)$ are observed during the
dropouts, see Fig.~\ref{fig:LFF-example}. This is very intuitive as $N(t)>0$ corresponds to the light amplification regime of the laser without feedback.

\section{Chaotic Bursting\label{sub:Slowfast}}

In this section, we propose a description of a mechanism responsible
for LFFs in the LK model using singular geometric perturbation theory.
We describe how LFFs can be viewed as a sequence of slow and fast
solution segments obeying a reduced set of equations. In particular,
we reveal how the increasing values of $N(t)$ lead to the dropout and
how the solution bursts off the non-lasing state $E(t)=0$. In addition,
we propose an averaged system of ODEs describing the intermediate
fast oscillations between the dropouts events for small $\varepsilon$. 

At first, we go through the stability analysis of the off-state $E(t)=0$.
Secondly, we investigate the limit $\varepsilon\to0$ and appeal to
invariant manifolds theory to provide a geometric understanding of
LFF (and RPP) events. As we intent to get across the main idea and
provide the underlying mechanism, we leave out the technical details.
We refer to Refs.~\onlinecite{Bates1998,Bates2000} for the theory
of invariant manifolds for semiflows. Via averaging, we obtain a related
ODE system and perform a slow-fast analysis of this system. We show
how the fast time-scale oscillations result from the dynamics residing
close to eigen-modes of the equilibrium $E(t)=0$ as long as it is stable
(in this average sense) and ''hop'' to another when it becomes unstable,
thereby revealing the mode-hopping mechanism.

\vspace{-0.1cm}
\subsection{Stability of the off-state}

\label{sub:laser-off} The linear stability of the steady state $(E(t),N(t))=(0,J)$
can be obtained from the corresponding linearized system 
\begin{eqnarray}
E^{\prime}\left(t\right) & = & \left(1+i\alpha\right)nE\left(t\right)+\eta e^{-i\phi}E\left(t-\tau\right),\label{eq:Elin}\\
N^{\prime}\left(t\right) & = & -\varepsilon N,\label{eq:Nlin}
\end{eqnarray}
where $n=J$ is the equilibrium value of $N(t)$. We use a new notation
$n$ instead of $J$ for the equilibrium value of $N$, since the
results obtained in this section will be useful also in the subsequent
Sec.~\ref{sub:Slowfast-DDE}, where $N(t)$ will admit also other
values different from $J$. 

Equations~\eqref{eq:Elin}\textemdash \eqref{eq:Nlin} are decoupled
and, in $N$-direction, the eigenvalue is $\lambda=-\varepsilon$.
The remaining spectrum consists of eigenvalues $\lambda$, which are
solutions of the characteristic equation 
\begin{equation}
-\lambda+\left(1+i\alpha\right)n+\eta e^{-i\phi-\tau\lambda}=0\label{eq:Efastchar}
\end{equation}
obtained by inserting the ansatz $E(t)=E(0)e^{\lambda t}$ into Eq.~\eqref{eq:Elin}.
The equilibrium is stable, if all eigenvalues have negative real parts. 

Firstly, let us note that the equilibrium is stable for $n<-\eta$.
This can be easily seen from the real part of the characteristic equation
\eqref{eq:Efastchar}
\[
\mbox{Re}\lambda=n+\eta e^{-\tau\mbox{Re}\lambda}\cos(\phi+\tau\mbox{Im}\lambda),
\]
which can be estimated as $\mbox{Re}\lambda\le n+\eta$ for all $\lambda$
such that $\mbox{Re}\lambda\ge0$ and $\eta>0$. Note that, generically,
$\lambda=0$ is not an eigenvalue, since $\lambda=0$ implies $\phi=\arctan\alpha+2k\pi,~k\in\mathbb{Z},$
which is a special case not considered here. 

\begin{figure}
\noindent %
\begin{minipage}[t]{.5\linewidth}%
\includegraphics[width=1\linewidth]{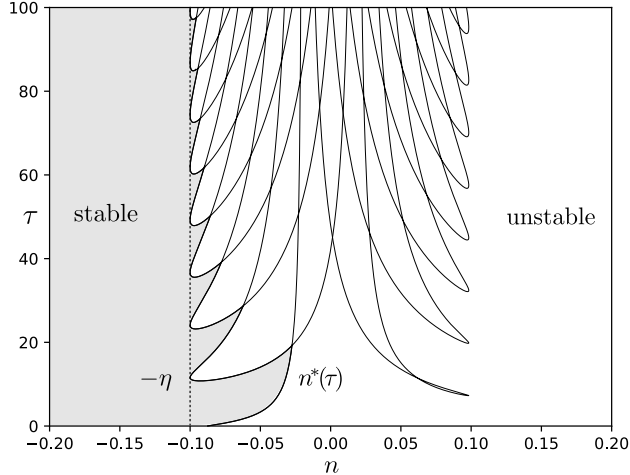}%
\end{minipage}\caption{Stability of the off-state $E(t)=0$. The figure shows the Hopf bifurcation
curves in the parameter plane $(n,\tau)$, $n\in[-0.2,0.2]$, $\tau\in[0,100]$
for the fixed parameter values $\alpha=5,~\eta=0.1,\phi=0.5$, and
$\varepsilon$ arbitrary (stability does not depend on $\varepsilon>0$).
Here, $n$ is the equilibrium value of the $N$-variable. $E(t)$
is stable independently on $\tau$, when $n<-\eta=-0.1$. As $n$
increases one pair of eigenvalues crosses the imaginary axis at each
Hopf bifurcation curve.\label{fig:ZeroStab}}
\end{figure}

Therefore, the equilibrium may destabilize via a bifurcation of Hopf-type,
where a purely imaginary eigenvalue $\lambda=i\omega$ crosses the
imaginary axis and gives rise to an ECM solution with period $2\pi/\omega$.
These Hopf bifurcation curves are shown in Fig.~\ref{fig:ZeroStab}.
The diagram can be interpreted as follows. First, for every fixed
$n$ there are values $\tau_{0}(n),\tau_{1}(n),\dots$ such that a
purely imaginary eigenvalue crosses the imaginary axis. Secondly,
for every fixed $\tau$, there are $n_{0}=n_{0}(\tau,\omega_{0}),~n_{1}=n_{1}(\tau,\omega_{1}),\dots$
such that there exist ECMs with periods $2\pi/\omega_{0},~2\pi/\omega_{1},\dots~$,
if $n_{0}=n_{0}(\tau,\omega_{0}),~n_{1}=n_{1}(\tau,\omega_{1}),\dots<J$.
Fig.~\ref{fig:ZeroStab} shows the stability
region of the off-state and illustrates that the number of ECMs grows proportionally
with the time-delay. 

As a next step, we discuss how large the critical eigenvalues are. 
More specifically, we are interested in
scaling properties of the eigenvalues for large delay. Accordingly
to the theory in Ref.~\onlinecite{Lichtner2011} (see also reviews
in Refs.~\onlinecite{YanchukGiacomelli2017,Wolfrum2011}), the
eigenvalues scale generically either as $1/\tau$ for large $\tau$
(weak instability) or as $\mathcal{O}(1)$ (strong instability), similarly
to the scaling of LEs for weak and, respectively, strong chaos discussed
in Sec.~\ref{sec:LE}. Both parts of the spectrum can be calculated.
We omit here the details of the straightforward calculations (see
e.g. Refs.~\onlinecite{Lichtner2011,YanchukGiacomelli2017,Wolfrum2011}),
and present the results. In particular, the strong spectrum is given
by 
\begin{equation}
\lambda_{\mbox{strong}}\approx(1+i\alpha)n,\quad\mbox{for}\quad n>0\label{eq:strongrate}
\end{equation}
and the weak spectrum can be approximated by the continuous curve
\[
\lambda_{\mbox{weak}}(\omega)=\frac{1}{\tau}\gamma(\omega)+i\omega=-\frac{1}{2\tau}\ln\frac{\left(\omega-\alpha n\right)^{2}+n^{2}}{\eta^{2}}+i\omega,
\]
$\omega\in\mathbb{R},$ as $\tau\to\infty$. If $n<0$, the strong
spectrum is absent and there is no positive eigenvalue with $\mbox{Re}\left(\lambda\right)=\mathcal{O}(1)$
as $\tau\to\infty$. Then, the largest real parts of eigenvalues are
approximately given by 
\begin{equation}
\max_{\omega}\mbox{Re}(\lambda)\approx\max_{\omega}\frac{\gamma(\omega)}{\tau}=\frac{\gamma(\alpha n)}{\tau}=\frac{1}{\tau}\ln\left|\frac{\eta}{n}\right|.\label{eq:weakrate}
\end{equation}
 Therefore, for large values of $\tau$ and $-\eta<n<0$, $E(t)=0$ is
so-called weakly unstable with the most unstable eigen-mode $e^{i\alpha nt}$.
If $n>0$, we call $E(t)=0$ strongly unstable, with most unstable eigen-mode
$e^{\left(1+i\alpha\right)nt}$.

\vspace{-0.1cm}
\subsection{Direct slow-fast analysis \label{sub:Slowfast-DDE}}

In this section, we present a mechanism for the recurrent appearance
of dropouts during LFFs. For this we use a direct slow-fast analysis
and smallness of parameter $\varepsilon$. We split the description
into two parts: Part 1 describes the dynamics close to the off-state
and the mechanisms that bound the orbit close to the off-state for
certain time and then leads to the amplitude increase, i.e. repelling
from the off-state. Part 2 describes the return mechanism for solutions
with large amplitude. The LFF recurrent mechanism then follows from
the combination of these two ingredients.

\paragraph*{Part 1. Dynamics close to the off-state, repelling mechanism. }\hspace{-.5cm}
Setting formally $\varepsilon=0$ in Eqs.~\eqref{eq:Edef}\textendash \eqref{eq:Ndef}, we obtain 
\begin{eqnarray}
E^{\prime}(t) & = & \left(1+i\alpha\right)N(t)E(t)+\eta e^{i\phi}E(t-\tau),\label{eq:Edef-2}\\
N^{\prime}(t) & = & 0,\label{eq:Ndef-2}
\end{eqnarray}
hence, one can consider Eq.~\eqref{eq:Edef-2} in a layer given by
a fixed value $n=N\left(0\right)$. As a result, we obtain Eq.~\eqref{eq:Elin}, which was studied in the previous section in detail.
In particular, $E(t)=0$ is an equilibrium for every $n\in\mathbb{R}$.
In other words, system with $\varepsilon=0$ possesses the line of
equilibria $M=\{(E,N)=(0,n)\,|\,n\in\mathbb{R}\}$.

The invariant line $M$ (or any connected compact subset of it) is
called normally hyperbolic, if the characteristic equation \eqref{eq:Efastchar}
has no solutions $\lambda=\lambda(n)$ with zero real part. Normal
hyperbolicity ensures persistence of $M$ under small perturbations \cite{Bates1998,Bates2000}.
As follows from Sec.~\ref{sub:laser-off}, for every $\tau$ there
exists $n^{\ast}(\tau)\geq-\eta$ such that for $n<n^{\ast}(\tau)$,
the line of equilibria parametrized by $n$ is normally exponentially
stable with uniform bounds away from $n^{\ast}(\tau)$ (and unstable
otherwise), see Fig.~\ref{fig:ZeroStab}, where $n^{*}(\tau)$ is
the stability boundary.

\begin{figure}
\noindent %
\begin{minipage}[t]{.5\linewidth}%
\includegraphics[width=1\linewidth]{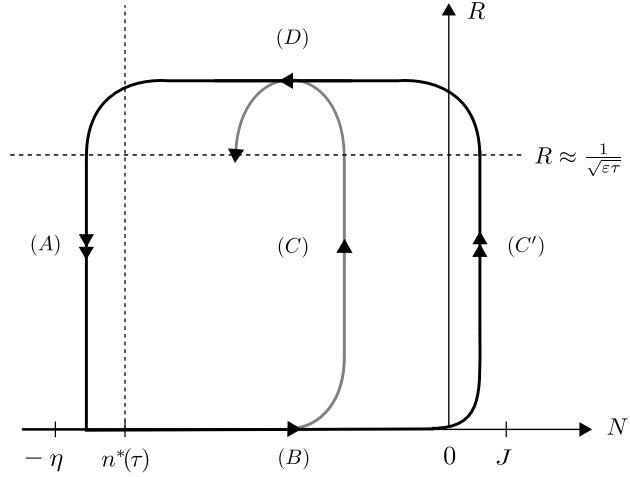}%
\end{minipage}\caption{\label{fig:Relax-1}Slow-fast solution segments for $\varepsilon$
small. Fast segment (A): for $N(t)<n^{\ast}(\tau)$ the solution is
attracted by the zero solution $R(t)=0$. Slow segment (B): for $N(t)<n^{\ast}(\tau)$
the solution stays close to $R(t)=0$ and $N(t)$ slowly increases
towards $N(t)=J$. Segment (C): for $n^{\ast}(\tau)<N(t)<0$ and $\tau$
large, $R(t)$ increases slowly for large $\tau$ with the rate of
increase bounded by $-\ln|n/\eta|/(2\tau)$. $R^{\prime}(t)=0$ in
each layer containing an ECM. Fast segment (C'): For $N(t)>0$ the
amplitude increases fast. Return flow (D): If $R(t)\geq1/\sqrt{\tau\varepsilon}$,
$N(t)$ is strictly decreasing for the considered parameters. See
Sec.~\ref{sub:Slowfast-DDE} for details on black segments and Sec.~\ref{sub:Slowfast-ODE}
for details on gray segments.}
\end{figure}

In order to reveal the slow dynamics on $M$, we rescale time as $\tilde{t}=\varepsilon t$
in Eqs.~\eqref{eq:Edef}\textendash \eqref{eq:Ndef} and obtain 
\begin{eqnarray}
\varepsilon\dot{E}(\tilde{t}) & = & \left(1+i\alpha\right)N(\tilde{t})E(\tilde{t})+\eta e^{i\phi}E(\tilde{t}-\varepsilon\tau),\label{eq:Edef-3}\\
\dot{N}(\tilde{t}) & = & \left[J-N(\tilde{t})-\left(1+2N(\tilde{t})\right)|E(\tilde{t})|^{2}\right],\label{eq:Ndef-3}
\end{eqnarray}
where the dot denotes differentiation with respect to the new slow
time. Considering the limiting case $\varepsilon=0$ and $E(t)=0$, it
is easy to see that the slow flow on $M$ is given by the ODE $\dot{N}(\tilde{t})=J-N(\tilde{t}),$
with the stable equilibrium $N(\tilde{t})=J$. The above facts allow the description of the dynamical picture for small $\varepsilon$. We refer the reader to Fig.~\ref{fig:Relax-1} for a visualization of the following paragraph. 

Assume that at some point of time $N(t)=n<-\eta$ holds. Then, as
follows from Sec.~\ref{sub:laser-off}, the zero equilibrium of the
layer equation \eqref{eq:Edef-2} is globally stable and, hence, the
solution converges to the off-state $E(t)=0$, i.e. the point ($E(t)=0,N(t)=n$)
on the invariant line $M$, see segment (A) in Fig.~\ref{fig:Relax-1}. 

On the manifold $M$, the solution slowly (accordingly to the timescale
$\tilde{t}=\varepsilon t$) moves towards the point $N(\tilde{t})=J$, $E(t)=0$
which is the stable equilibrium within $M$, see segment (B) in Fig.~\ref{fig:Relax-1}.
In the case, when $J<n^{*}(\tau)$, the solution just converges to the
stable off-state. However, for the LFF case, it holds $J>n^{*}(\tau)$.
Hence, as the value of $N(\tilde{t})$ increases, $M$ loses
normal stability at $n^{\ast}(\tau)$, and it causes $R(t)=|E(t)|$
to increase with the approximate rate $\frac{1}{\tau}\ln(\left|\eta/N(\tilde{t})\right|)$
(see Eq.~\eqref{eq:weakrate}) when $N(\tilde{t})<0$, see segment
(C) in Fig.~\ref{fig:Relax-1}. For large $\tau$, this rate of increase
can be very slow, and, in particular, the interplay between $\varepsilon$
(speed of the motion along $M$) and $1/\tau$ (repelling rate) can
play a decisive role in determining the time the solution spends in the
vicinity of the off-state. Moreover, if $J>0$, then $N(\tilde{t})$
can become positive, which leads to even faster repelling rate, see
Eq.~\eqref{eq:strongrate}, that is independent of $\tau$, see segment
(C') in Fig.~\ref{fig:Relax-1}. The above mechanism describes the
attraction by and repelling from the off-state, which determines the
dropout event

\paragraph*{Part 2. Return mechanism for large-amplitude solutions.}
\hspace{-.3cm} Note that for values 
\[
R(t)>\sqrt{\frac{J-N(t)}{1+2N(t)}}
\]
it holds $N'(t)<0$ and $N(t)$ is strictly decreasing. This leads
to a return to smaller values of $N(t)$. Indeed, as soon as $N<n^{*}(\tau)$,
the equilibrium in the layer equation of $E(t)$ becomes again stable
and the solution converges back to the off-state, see segments (D) and
(A) in Fig.~\ref{fig:Relax-1}.

The return mechanism can be explained more rigorously for the case
when the amplitude $R$ becomes large. For instance, let us assume
that $R\left(t\right)\approx1/\sqrt{\varepsilon\tau},$ then the rescaled
variable $\tilde{E}(t)=\sqrt{\varepsilon\tau}E(t)$ can be introduced,
and the resulting system has the form of the perturbed equation 
\begin{eqnarray}
\tilde{E}^{\prime}\left(t\right) & = & \left(1+i\alpha\right)N\left(t\right)\tilde{E}\left(t\right)+\eta e^{i\phi}\tilde{E}\left(t-\tau\right),\label{eq:Edef-1}\\
N^{\prime}\left(t\right) & = & -\frac{1}{\tau}\left(1+2N\left(t\right)\right)\left|\tilde{E}\left(t\right)\right|^{2}+\varepsilon\left(J-N\left(t\right)\right).\qquad \label{eq:Ndef-1}
\end{eqnarray}
Since $J-N\sim\eta$, the term $\varepsilon\left(J-N\left(t\right)\right)\sim\varepsilon\eta$
is smaller than $1/\tau$ for the parameters considered. Hence, by
neglecting this term, it follows that $N^{\prime}\left(t\right)\leq0$
and $N\left(t\right)\to-1/2<-\eta$ as $t\to\infty$. As a result,
$\tilde{E}\left(t\right)\to0$ as $t\to\infty$. In particular, $\tilde{E}(t)$
decreases until the rescaling is not justified anymore with some value
$N\left(t\right)=n$. If again $n<n^{\text{\ensuremath{\ast}}}(\tau)$,
$E(t)=0$ is stable in this layer and by the fast flow, the solution
converges to $E(t)=0$ and we obtain the recurrent LFF behavior combining
the mechanisms described in part 1 and part 2 above. Solutions that
enter the regime of small amplitude for values $N(t)=n>n^{\text{\ensuremath{\ast}}}(\tau)$
however, do not fit this simple description. The equilibrium here
is of (high-dimensional) saddle type, so that very complicated dynamics
may arise.

\vspace{-0.1cm}
\subsection{Averaging the fast flow}

\label{sub:averaging} In order to describe the fast oscillatory behavior
of the solution when $-\eta<n^{\ast}(\tau)<N(t)<0$, for $\eta$
small, in this section, we propose an averaging method leading to
a system of ODEs. We define the exponential change rate of $E(t)$
between two subsequent delay intervals as 
\begin{equation}
\Lambda\left(t\right):=\frac{1}{\tau}\mbox{Log}\left(\frac{E\left(t\right)}{E\left(t-\tau\right)}\right),\label{eq:Lambdat}
\end{equation}
such that $E(t)=E(t-\tau)e^{\tau\Lambda(t)}$ and denote $\Gamma(t):=\mbox{Re}(\Lambda(t))$.
Straightforward computation reveals that 
\[
\limsup_{L\to\infty}\frac{1}{L}\sum_{l=0}^{L}\Gamma\left(t_{0}+l\right)=\lambda_{E},
\]
for all $t_{0}\geq0$, where $\lambda_{E}$ is the LE of the E-variable,
determined from of Eq.~\eqref{eq:Edef} only. Therefore $\Gamma\left(t\right)$
can be interpreted as a finite time LE \cite{Kanno2014} restricted
to the $E$-component of the LK-system. Clearly, positivity of this
LE implies positivity of a LE in the full system. $\Lambda(t)$ satisfies 
\begin{equation}
\tau\Lambda^{\prime}\left(t\right)=\frac{E^{\prime}(t)}{E(t)}-\frac{E^{\prime}(t-\tau)}{E(t-\tau)},\label{eq:FTLEevol}
\end{equation}
 where
\begin{equation}
\frac{E^{\prime}(t)}{E(t)}=(1+i\alpha)N(t)+\eta e^{i\phi-\tau\Lambda(t)}.\label{eq:FTLE-aux}
\end{equation}
As $|N(t)|<\eta$ and, moreover, suggested by numerics, the assumption
$\tau\mbox{Re}(\Lambda(t))\sim1$ (see, e.g. Fig.~\ref{fig:LFF-example}(b))
leads to the estimate
\[
\left|\frac{E^{\prime}(t)}{E(t)}\right|\leq C\eta.
\]
As a result, the right hand side of Eq.~\eqref{eq:FTLEevol} is bounded
by $2C\eta$, where $C$ is independent of $\eta$. Hence, we can
consider $h(t)={E^{\prime}(t-\tau)}/{E(t-\tau)}\sim\eta$ as a time-dependent
small perturbation and replace it by its average over length $\tau$
\begin{equation}
\frac{1}{\tau}\int\limits _{0}^{\tau}h(t+s)ds=\frac{1}{\tau}\int\limits _{0}^{\tau}\frac{E^{\prime}\left(t+s-\tau\right)}{E\left(t+s-\tau\right)}ds=\Lambda\left(t\right).\label{eq:subst}
\end{equation}
to obtain the following ODE 
\begin{equation}
\tau\Lambda^{\prime}\left(t\right)=\frac{E^{\prime}(t)}{E(t)}-\Lambda\left(t\right).\label{eq:FTLEevol-av}
\end{equation}
Using this ODE, one can express $R(t)$ as a function of $\Gamma(t)$.
For this, by integrating \eqref{eq:FTLEevol-av} and taking the absolute
value, we obtain 
\begin{equation}
R(t)=R(0)\exp\left[\tau\left(\Gamma(t)-\Gamma(0)\right)+\int_{0}^{t}\Gamma(s)ds\right].\label{eq:R}
\end{equation}
 Now, substituting \eqref{eq:FTLE-aux} into \eqref{eq:FTLEevol-av},
and combining with \eqref{eq:Ndef}, we derive the following system
for $\Lambda(t)$ and $N(t)$ 
\begin{eqnarray}
\tau\Lambda^{\prime}\left(t\right) & = & \left(1+i\alpha\right)N\left(t\right)+\eta e^{i\phi}e^{-\tau\Lambda\left(t\right)}-\Lambda\left(t\right),\label{eq:AFTLE}\\
N^{\prime}\left(t\right) & = & \varepsilon\left[J-N\left(t\right)-\left(1+2N\left(t\right)\right)R^{2}\left(t\right)\right],\label{eq:AFTLN}
\end{eqnarray}
where $R(t)$ is given by \eqref{eq:R}. The equation for $R$ can
be also written in a differential form 
\begin{equation}
R^{\prime}\left(t\right)=\left(N\left(t\right)+\eta e^{-\tau\Gamma\left(t\right)}\cos\left(\tau\Omega\left(t\right)-\phi\right)\right)R\left(t\right),\label{eq:AFTLR}
\end{equation}
thus, completing the system \eqref{eq:AFTLE}\textendash \eqref{eq:AFTLR}.
The right hand side of each equation satisfies the following order
estimates $\Lambda'\lesssim\eta/\tau$, $N'\lesssim\varepsilon$,
and $R'\lesssim\eta$. In our case, all three estimations are small,
which are the necessary conditions for the application of the averaging,
and which ensure a certain closeness of the solutions of the averaged
system \eqref{eq:AFTLE}\textendash \eqref{eq:AFTLR} and the corresponding
quantities $\Lambda(t)$, $R(t)$, and $N(t)$ obtained from the solutions
$E(t)$ and $N(t)$ of the original LK system using the relations
\eqref{eq:FTLEevol-av} and $R(t)=|E(t)|$. The in-depth analysis
of the closeness of the solutions to averaged system and the LK system
is very technical and out of the scope of this manuscript. Instead,
in the next section, we study the averaged system for small $\varepsilon$.

\vspace{-0.1cm}
\subsection{Slow-fast analysis of the averaged system \label{sub:Slowfast-ODE}}

We start with the limit $\varepsilon=0$, where the equations in a
layer $N(t)=n$ of the averaged system \eqref{eq:AFTLE}\textendash \eqref{eq:AFTLR}
are given by 
\begin{eqnarray}
\tau\Lambda^{\prime}\left(t\right) & = & \left(1+i\alpha\right)n+\eta e^{i\phi}e^{-\tau\Lambda\left(t\right)}-\Lambda\left(t\right),\label{eq:AFTLE-1}\\
R^{\prime}\left(t\right) & = & \left(n+\eta e^{-\tau\Gamma\left(t\right)}\cos\left(\tau\Omega\left(t\right)-\phi\right)\right)R\left(t\right).\quad\label{eq:AFTLR-1}
\end{eqnarray}
Note that Eq.~\eqref{eq:AFTLE-1} is decoupled and does not depend
on $R(t)$. One of the remarkable features of this system is that
its equilibria $\Lambda(t)=\lambda$ satisfy Eq.~\eqref{eq:Efastchar},
that is, the equilibria coincide with the eigenvalues of the characteristic
equation for the off-state. Generically, there are countably many
equilibria, since \eqref{eq:Efastchar} is a quasi-polynomial. For
a detailed analysis of Eq.~\eqref{eq:Efastchar}, see Sec.~\ref{sub:laser-off}.

We linearize Eqs.~\eqref{eq:AFTLE-1}\textendash \eqref{eq:AFTLR-1}
at such an equilibrium, where $\Lambda(t)=\lambda$ and $R(t)=0$.
As Eq.~\eqref{eq:AFTLE-1} is decoupled, it is easy to check that
the corresponding eigenvalues are given by $\mu_{1}=(1+i\alpha)n-\lambda-1/\tau$
and $\mu_{2}=\gamma$, where $\gamma=\mbox{Re}(\lambda)$. Thus, at
$\gamma=0$, a simple, real eigenvalue crosses the imaginary axis.
Moreover, $\mu_{1}$ crosses the imaginary axis at $\gamma_{H}=n-1/\tau$.
To sum up, all equilibria $\lambda=\gamma+i\omega$ satisfying $\gamma_{H}<\gamma<0$
are stable, and unstable for $\gamma>0$ or $\gamma<\gamma_{H}$.

It is interesting that these equilibria correspond to the eigen-modes
$E(t)=e^{\lambda t}$ of Eq.~\eqref{eq:Elin}. In particular, we
have shown that eigen-modes, which cause the amplitude of the electric
field to increase $(\gamma>0)$ are unstable in the average sense
above.

Similarly to Sec.~\ref{sub:Slowfast-DDE}, these results hold for
an arbitrary $n$, such that we obtain countably many curves of equilibria
$\lambda=\lambda(n)$ across layers, and each of these curves is normally
exponentially stable if $n-1/\tau<\mbox{Re}\left(\lambda\left(n\right)\right)<0$,
with uniform bounds away from $n-1/\tau$ and 0. Hence, each of these
normally exponentially stable pieces of the curves persist as invariant
manifolds for $\varepsilon>0$. In order to obtain the slow flow on
these curves, we rescale time as $\tilde{t}=\varepsilon t$ and consider
the limit $\varepsilon=0$, to find that in the limit $\dot{N}(\tilde{t})=J-N(\tilde{t})$,
see similar calculations in Sec.~\ref{sub:Slowfast-DDE}. Hence, the solution
follows the slow manifold close to $(\Lambda=\lambda(n),R=0,N=n)$
with increasing $n$ until it destabilizes or reaches $N=J$. 

We remark, that Eqs.~\eqref{eq:AFTLE-1}\textendash \eqref{eq:AFTLR-1}
attain a second set of equilibrium solutions $(\Lambda=i\omega,R=\mbox{const},N=n)$, where the value of $R$
is arbitrary and $\omega$, $N(t)$ satisfy Eq.~\eqref{eq:Efasthopf}. These correspond to the invariant co-rotating frames in
the limit $\varepsilon=0$ for the original Eqs.~\eqref{eq:Edef}\textendash \eqref{eq:Ndef},
each containing one ECM. Note however, that these curves are not normally
hyperbolic and hence, they do not persist for $\varepsilon>0$.

\begin{figure}
\noindent %
\begin{minipage}[t]{.5\linewidth}%
\includegraphics[width=1\linewidth]{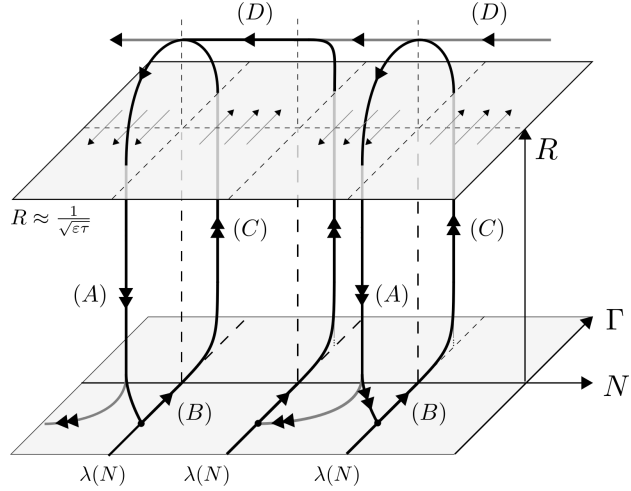}%
\end{minipage}\caption{Slow-fast solution segments of averaged system \eqref{eq:AFTLE}\textendash \eqref{eq:AFTLN}
for $n^{\ast}(\tau)<N(t)<0$ (compare to segment (C) in Fig.~\ref{fig:Relax-1}). Fast segment (A): for $N(t)-1/\tau<\Gamma(t)<0$ ($N(t)$ is sufficiently
small here) the solution is attracted fast by a branch $\lambda(N(t))$
of eigenvalues of $R(t)=0$. Slow segment (B): for $\Gamma(t)<0$
the solution stays close to $R(t)=0,~\lambda(N(t))$ and $N(t)$ slowly
increases towards $N(t)=J>0$. Fast segment (C): for $\Gamma(t)>0$,
$R(t)$ increases fast. Return flow (D): If $R(t)\geq1/\sqrt{\varepsilon}$,
$N(t)$ is strictly decreasing the parameters considered. If $\Gamma(t)$
is again negative, the solution is again attracted by a branch of
eigenvalues (not necessarily the same). See Sec.~\ref{sub:Slowfast}
for details.\label{fig:Relax-2}}
\end{figure}

In the following paragraph we describe the mechanism of slow-fast oscillations in system \eqref{eq:AFTLE}--\eqref{eq:AFTLR}
using similar arguments as in Sec.~\ref{sub:Slowfast-DDE}. Assume
that at some point of time $N(t)-1/\tau<\Gamma(t)<0$, with $N(t)=n$
holds. Then, the solution converges fast to a stable equilibrium $\lambda(n)$,
$E(t)=0$, i.e. a point on the invariant curve $\left\{ n\in\mathbb{\mathbb{R}}|\Lambda(t)=\lambda(n),E(t)=0,N(t)=n\right\} $,
see segment (A) in Fig.~\ref{fig:Relax-2}. $\lambda\left(N(\tilde{t})\right)$
then evolves according to the slow flow on $\lambda(n)$, see segment (B) Fig.~\ref{fig:Relax-2}.
If, as $N(\tilde{t})$ increases by the slow flow along $\lambda\left(N(\tilde{t})\right)$,
$\lambda\left(N(\tilde{t})\right)$ destabilizes, $R\left(t\right)$
increases fast, see segment (C) in Fig.~\ref{fig:Relax-2}. There are no further
equilibria in a layer, so we construct a return flow similar to
Sec.~\ref{sub:Slowfast-DDE}. As $R\left(t\right)$ increases, we
can rescale $\tilde{R}(t):=R(t)/\sqrt{\varepsilon\tau}$. When $\tilde{R}\left(t\right)=\mathcal{O}\left(1\right),$
the corresponding rescaled system takes the form 
\begin{eqnarray}
\tau\Lambda^{\prime}\left(t\right) & = & \left(1+i\alpha\right)N\left(t\right)+\eta e^{i\phi-\tau\Lambda\left(t\right)}-\Lambda\left(t\right),\label{eq:AFTLE-2}\\
\tilde{R}^{\prime}\left(t\right) & = & \left(N\left(t\right)+\eta e^{-\tau\Gamma\left(t\right)}\cos\left(\tau\Omega\left(t\right)-\phi\right)\right)\tilde{R}\left(t\right),\,\,\,\,\,\,\,\,\,\,\,\,\,\,\label{eq:AFTLR-2}\\
N^{\prime}\left(t\right) & = & -\frac{1}{\tau}\left(1+2N\left(t\right)\right)\tilde{R}^{2}\left(t\right)+\varepsilon\left[J-N\left(t\right)\right].\label{eq:AFTLN-1}
\end{eqnarray}
We use further similar arguments as in Sec.~\ref{sub:Slowfast-DDE},
we remark that the term $\varepsilon\left[J-N\left(t\right)\right]\sim\varepsilon\eta$
is much smaller than $1/\tau$ and, hence, it can be neglected. Strikingly,
we again find curves of eigenvalues $\Lambda(t)=\lambda(n),~R(t)=0,~N(t)=n$,
as the equilibrium value of $N(t)$ is arbitrary, identical to the
non-rescaled case. Moreover, their stability is identical to the stability
of the critical curves of Eqs.~\eqref{eq:AFTLE}\textendash \eqref{eq:AFTLN}
(with a zero eigenvalue along each curve). It is apparent however,
that if $N(t)>-1/2$ and not stationary, $N\left(t\right)$ is decreasing,
providing us with a return flow as indicated for segment (D) in Fig.~\ref{fig:Relax-2}.
In particular, as $\Lambda(t)=\lambda(N(t))$ solves Eq.~\eqref{eq:AFTLE-2},
the sign of $\mbox{Re}(\Lambda(t))$ must change, when crossing $\lambda(N(t))$,
such that we can expect that $\mbox{Re}(\Lambda(t))$ is also negative
at some point and the solution again converges to one of the curves
$\lambda(n)$. This geometric viewpoint offers great insight into
the intermediate behavior of solutions, when $-\eta<n^{\ast}(\tau)<N(t)<0$
and fills in the gap outlined in Sec.~\ref{sub:Slowfast-DDE}.

\section{Discussion}

In this manuscript, we numerically investigated the parameter
regions, where LFFs and RPPs occur. 
We have shown that $\varepsilon\tau$ is an important quantity for the solution to exhibit LFFs. In fact, multiple time scale analysis suggests that $\varepsilon\tau<1/\eta$ is a necessary condition for LFFs to occur.
In particular, we numerically observe that LFFs cease to exist for large time-delays $\tau>1/\varepsilon\eta$, if all other parameters are fixed.
Additionally, we studied the spectrum of these solutions and conclude that LLFs are observed in the regime of weak chaos, at least for the considered parameter values. 
In the second part of the paper, we use a multi scale approach to
characterize such solutions as bursting slow-fast oscillations. In
particular, using singular geometric perturbation theory for delay
differential equations, we have shown that within a dropout event
the system can be reduced to a slow flow towards $N(t)=J$ along the
zero solution $E(t)=0$ and that the amplitude builds up after the dropout
is due to a weak instability for $n^{\ast}(\tau)<N(t)<0$ or strong
instability for $N(t)>0$. Here $n^{\ast}(\tau)$ is the destabilization
threshold for the off-state, which is explicitly computed. 
In order to describe the weakly chaotic solution for $E(t)>0$, we
have imposed a finite-time averaging technique and investigated the
corresponding system of ordinary differential equations. This approach
can be considered as a generalization of reduced models that have
already been applied in the analysis of the LK system \cite{Lenstra1991, Huyet2000}
and similar steps have been taken in the statistical theory of phase
oscillators \cite{DHuys2014}. Our averaging method relies on the
smallness of the parameters $\varepsilon$, $1/\tau$, and $\eta$,
which are very natural for the LK system. Our analysis suggests that
this method is applicable to a larger class of weakly chaotic solutions
of singularly perturbed delay differential equations, where phenomena
similar to LFFs exists. See Ref.~\onlinecite{Pieroux2003}
for an example in the context of the LK model.
\begin{acknowledgments}
\noindent This paper was developed within the scope of the IRTG 1740/
TRP 2015/50122-0, funded by the DFG/ FAPESP. The authors would like
to thank Giovanni Giacomelli, Matthias Wolfrum and Jaap Eldering for
fruitful discussion. 
\end{acknowledgments}

%

\end{document}